\documentclass[a4paper]{jpconf}
\usepackage{graphicx}
\begin{document}

\title{Parton promenade into the nucleon}

\author{Eric Voutier}

\address{Laboratoire de Physique Subatomique et de Cosmologie, CNRS/IN2P3,
Universit\'e Joseph Fourier, INP, 53 rue des Martyrs, 38026 Grenoble cedex,
France}

\ead{voutier@lpsc.in2p3.fr}

\begin{abstract}
Generalized parton distributions (GPDs) offer a comprehensive picture of the 
nucleon struture and dynamics and provide a link between microscopic and 
macroscopic properties of the nucleon. These quantities, which can be 
interpreted as the transverse distribution of partons carrying a certain 
longitudinal momentum fraction of the nucleon, can be accessed in deep 
exclusive processes. This lecture reviews the main features of the nucleon 
structure as obtained from elastic and inelastic lepton scatterings and 
unified in the context of the GPDs framework. Particular emphasis is put on 
the experimental methods to access these distributions and the today
experimental status.
\end{abstract}

\section{Introduction}

The nucleon, this most singular object of nuclear physics whose mass is quite 
larger than the mass of its constituents, did resist since several decades to 
theoretical and experimental investigations. It is only recently that a
comprehensive picture of the nucleon structure started to develop within the 
framework of the generalized parton distributions (GPDs)~\cite{{Mul94},{Rad97}}. 
These distributions parametrize the partonic structure of the nucleon in terms 
of correlations between quarks, anti-quarks and gluons, and therefore contain 
information about the dynamics of this system. The power of this framework for 
the problem of the nucleon structure is certainly within the Mellin moments of 
the GPDs~\cite{Die03} which provide a natural link between microscopic and 
macroscopic properties of the nucleon.

GPDs can be accessed in the Bjorken regime of deep exclusive
processes~\cite{{Ji98},{Col99}}, that is when the resolution power of the probe 
is large enough to resolve partons and when the momentum transfer to the 
nucleon is small enough to insure the separation of perturbative and 
non-perturbative scales. Pioneer measurements at HERMES~\cite{Air01} and 
CLAS~\cite{Ste01}, and recent experiments at the Jefferson 
Laboratory~\cite{{Mun06},{Gir08}} have established the relevance of the deeply 
virtual Compton scattering (DVCS) process for this studies. The real and 
imaginary parts of the DVCS amplitude are the quantities of interest that can 
be separated via beam charge, or beam polarization, or target polarization 
observables. 

This lecture reviews the main steps of the story of the nucleon structure and
its modern expression within the GPDs framework. The experimental methods to
access these distributions and the today experimental status are also 
discussed.

\section{From nucleon to partons...}

From an experimental point of view, the story of the nucleon structure started
in the fifties when deviations from the Mott cross section were observed in
elastic electron scattering~\cite{Hof55}, meaning that the nucleon is not a 
pointlike object. The size of the nucleon is expressed by the so-called 
electromagnetic form factors which characterize the nucleon shape with respect 
to the electromagnetic interaction. This shape depends not-only on the 
resolution of the probe, controlled by the momentum transfer $Q^2$ to the 
nucleon, but also on the nature of the probe, this last feature being an early 
indication of the complexity of the nucleon structure. Within a non-relativistic 
framework, the electric (magnetic) form factor is the Fourier transform of the 
charge(current) density. Relativistic corrections to this picture modify this 
simple relationship but preserve this intimate link~\cite{Kel02}. As today,
there exists a significant enough knowledge of these form factors to question
the experimental disagreement between Rosenbluth and recoil polarization 
measurements~\cite{Per07}, as well as the role and magnitude of the two photons 
exchange mechanism~\cite{Gui03}.

At the end of the sixties, the nucleon revealed an unexpected behaviour in deep 
inelastic electron scattering (DIS). For excitation energies well beyond the
resonance region, the cross section was found to weakly depend on the momentum
transfer as compared to elastic scattering, indicating an interaction off a 
structureless object~\cite{Bre69}. This behaviour was later identified as the 
first evidence of the existence of partons. The cross section for these 
experiments depends on the variable $x_B$, the fraction of the nucleon 
longitudinal momentum carried by the partons, and can be expressed in 
terms of the probability to find in the nucleon a parton carrying a given 
longitudinal momentum. This feature led to extensive measurements of the 
momentum distributions of partons into nucleons, the so-called parton 
distributions whose statisfactory knowledge has now been obtained after 
thirty years of experimental efforts. The nucleon appears as a system of three
valence quarks globally equally sharing the nucleon momentum, and lying in a 
sea of quark-antiquark pairs and gluons.  

Soon after this discovery, it was realized that deep inelastic scattering of
polarized leptons off polarized targets allows to experimentally access the 
spin of the nucleon. Helicity conservation at the $\gamma^\star$-quark 
interaction vertex requires the $\gamma^\star$ coupling with opposite helicity
quarks. Therefore, changing the polarization of the nucleon allows to measure
alternatively the population of quarks with spin parallel or anti-parallel to
the nucleon. Within the naive parton model, the difference between these 
populations is a measurement of the nucleon spin. This picture turns out 
to be too rough when it was found that quarks carry about 20-30~\% of the total 
spin of the nucleon~\cite{{Ash88},{Ash89}}. In a more realistic partonic picture, 
the nucleon spin gets orbital and spin contributions of each constituent in a scale 
dependent manner, and the dynamics of the system expresses differently in the 
longitudinal (helicity) and transverse (transversity) directions yielding two 
spin sum rules~\cite{Bak04}. If a fair knowledge of quark helicity distributions 
has now been obtained, investigations of the gluon contribution and 
transversity distributions are only starting, and the orbital momentum 
contributions are essentially unknown. Overall, the spin of the nucleon 
remains by many respects a terra incognita.
    
GPDs procure the cement of the nucleon spin puzzle and more generally of the 
hadron structure.                                                                                                    

\section{... And conversely}

GPDs are universal non-perturbative objects entering the description of hard 
scattering processes and correspond to the amplitude for removing a parton of 
longitudinal momentum fraction $x+\xi$ and restoring it 
with momentum fraction $x-\xi$ (fig.~\ref{fig:hdbg}). In this process, 
the nucleon receives a four-momentum transfer $t=\Delta^2$ whose transverse 
component $\Delta_{\perp}$ is Fourier conjugate to the transverse position of 
partons. Consequently, GPDs can be interpreted as a distribution in the
transverse plane of partons carrying a certain longitudinal 
momentum~\cite{{Bur00},{Ral02},{Die02},{Bel02}}, constituting a 
femto-tomography of the nucleon.

At the leading twist, the partonic structure of the 
nucleon~\cite{{Die03},{Bel05}} is described by four quark helicity conserving 
and chiral even GPDs ($H^q, \widetilde{H}^q, E^q,\widetilde{E}^q$) and four 
quark helicity flipping and chiral odd GPDs ($H^q_T, \widetilde{H}^q_T, E^q_T,
\widetilde{E}^q_T$), together with eight similar gluon GPDs. In the forward 
limit ($t \to 0 , \xi \to 0$), the optical theorem links the $H$ GPDs to the 
usual density, helicity, and tranversity distributions measured in DIS. 
However, the $E$ GPDs, which involve a flip of the nucleon spin, do not have 
any DIS equivalent and then constitute a new piece of information about the 
nucleon structure. The first Mellin moments relate chiral even GPDs to form 
factors, as $E^q$ with the Pauli electromagnetic form factor 
\begin{equation}
\int_{-1}^{+1} dx \,\, E^q(Q^2,x,\xi,t) = F^q_2(t) \, ,
\end{equation}
and the second Mellin moments relate GPDs to the nucleon dynamics, 
particularly the total angular momentum carried by the partons, following 
Ji's sum 
rule~\cite{Ji97}  
\begin{equation}
J^q = \frac{1}{2} \int_{-1}^{+1} dx \,\, x \left[ H^q(Q^2,x,\xi,0) + E^q(Q^2,x,\xi,0) 
\right] \, .
\end{equation}
Here, $E$ is of particular interest since it is not constrained 
by any DIS limit and remains essentially unknown. Similar relations have been 
proposed which relate chiral odd GPDs to the transverse spin-flavor dipole 
moment and the correlation between quark spin and angular momentum in an 
unpolarized nucleon~\cite{Bur05}.

\section{Deep exclusive processes}

\begin{figure}[h]\centering
\includegraphics[width=59mm]{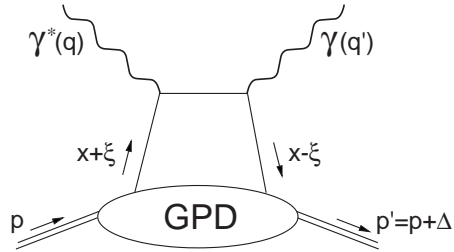}
\caption{Lowest order (QCD) amplitude for the virtual Compton process. The 
momentum four-vectors of the incident and scattered photon are $q$ and $q'$, 
respectively. The momentum four-vectors of the initial and final proton are 
$p$ and $p'$, with $\Delta$=$(p'-p)$=$(q-q')$. The DIS scaling variable is
$x_{\rm B}$=$Q^2/(2 p \cdot q)$ and the DVCS scaling variable is $\xi
$=$x_{\rm B}/(2-x_{\rm B})$.}
\label{fig:hdbg}
\end{figure}
DVCS, corresponding to the absorption of a virtual photon by a quark followed 
quasi-instantaneously by the emission of a real photon, is the simplest 
reaction to access GPDs. In the Bjorken regime, $-t \ll Q^2$ and $Q^2$ much 
larger than the quark confinement scale, the leading contribution to the 
reaction amplitude is represented by the so-called handbag diagram 
(fig.~\ref{fig:hdbg}) which figures the convolution of a known $\gamma^{\ast} 
q \to \gamma q$ hard scattering kernel with an unknown soft matrix element 
describing the partonic structure of the nucleon parametrized by GPDs. 
Consequently, GPDs enter the reaction cross section through a Compton form 
factor which involves an integral over the intermediate quark propagator and 
leads to a complex DVCS amplitude. 

In addition to the DVCS amplitude, the cross section for electroproduction of 
photons gets contributions from the Bethe-Heitler (BH) process where the real 
photon is emitted by the initial or final lepton, leading to~\cite{Die09} 
\begin{equation}
\sigma ( ep \rightarrow ep\gamma ) = \sigma_{BH} + \sigma_{DVCS} + P_l \, 
\widetilde{\sigma}_{DVCS} + e_l \, \sigma_{INT} + P_l e_l \, \widetilde{\sigma}_{INT}
\end{equation}
where the $\sigma$($\widetilde{\sigma}$)'s are even(odd) function of the 
out-of-plane angle between the leptonic and hadronic planes; $P_l$ and $e_l$ are 
the lepton polarization and charge, respectively. Though undistinguishable from 
DVCS, the BH cross section is known and exactly calculable from the 
electromagnetic form factors. The pure DVCS and interference contributions 
contain the information of interest, particularly $\sigma_{INT}$(
$\widetilde{\sigma}_{INT}$) is proportionnal to the real(imaginary) part of the 
DVCS amplitude. From the above relation, it is obvious that four different 
measurements combining beam charge and/or polarization allow to extract the 
four unknown contributions involving different Compton form factor combinations. 
Polarized target observables add other combinations with different sensitivity 
to the GPDs. For instance, $E$ can be accessed from the difference between 
polarized cross section off the neutron for opposite lepton helicities, and 
from the target spin asymmetry for a transversely polarized proton.

Deeply Virtual Meson Production (DVMP), where the real photon is replaced by a 
meson, is another channel to access GPDs additionnally providing an elegant 
flavor decomposition. In this case, the factorization of the cross section 
applies only to longitudinal virtual photons and the GPDs entering the Compton 
form factors are further convoluted with a meson distribution amplitude. The 
measurement of the angular distribution of the decay products of the vector 
mesons allows to extract the longitudinal part of the cross section and the 
longitudinal polarization of the vector mesons, assuming $s$-channel helicity 
conservation. Other reaction mechanisms, like the 2-gluon exchange from a $q 
{\bar q}$ fluctuation of the virtual photon, may contribute to the production 
process. Similarly to DVCS, polarization observables help to single-out the 
pure handbag contributions, particularly the production of longitudinal 
$\rho^0$ off a transversely polarized proton target is an observable very 
sensitive to $E$~\cite{Goe01}.    

\section{Experimental status}

The pioneering studies of the electro-production of photons at
DESY~\cite{{Air01},{Adl01}} and JLab~\cite{Ste01} did prove the existence of the 
DVCS mechanism by measuring sizeable beam spin asymmetries (BSA) in the valence 
region and significant cross sections in the gluon sector. Other limited studies 
showed the importance of the beam charge~\cite{Air07-1} and the target 
polarization observables~\cite{Che06}. A recent remarkable JLab result is the 
strong indication for scaling in the valence region at $Q^2$ as low as 2~GeV$^2$, 
and the observation of an unexpected DVCS amplitude magnitude at JLab 
energies~\cite{Mun06}. This early scaling is also supported by the $\varphi$ 
angular dependence of the BSA measured at JLab with an unprecedented accuracy 
over a wide kinematic range~\cite{Gir08}. In general, GPD based calculations 
provide a reasonable but unsatisfactory agreement with these data which turn 
out to be fairly reproduced by a more conventional Regge approach~\cite{Lag07}. 
The significance of this duality has not yet been resolved. 

In the meson sector, the experimental status with respect to GPDs remains 
controversial. Sizeable BSAs have been reported for exclusive $\pi^0$ 
electro-production at JLab energies~\cite{DeM08}, which suggest that both 
longitudinal and transverse amplitudes contribute to the process. On the one 
hand, this forbids a direct GPD based interpretation of the data, and on the 
other hand, a Regge based approach fails to reproduce them. The longitudinal 
cross section for the electro-production of longitudinally polarized $\rho^0$  
was recently measured at JLab~\cite{Mor08}. Standard GPD calculations, 
particularly successful at high energies, do not reproduce data in the valence 
region, while calculations based on hadronic degrees of freedom are in very 
good agreement over the complete energy range scanned by the world data. The 
observation that strongly modified GPDs allow for a partonic interpretation of 
these measurements raises the question whether current GPD parametrizations 
must be revisited or the existence of strong higher twist corrections percludes 
a GPD wise interpretation in the valence region.

Concerning the quest for the angular momentum, DVCS measurements off the
neutron~\cite{Maz07} and off a transversely polarized proton~\cite{Air08}, as well 
as longitudinal $\rho^0$ production off a transversely polarized 
proton~\cite{Air09} were reported. The interpretation of these data in terms of 
GPDs can lead to a constraint on the angular momentum of the $u$ and $d$ 
quarks~\cite{Ell06}. This constraint is however highly model dependent. It is 
the goal of the future experimental programs to provide enough data to map out 
the GPDs and allow for a model independent determination of the angular 
momentum of the partons.

\section{Conclusion}

Since a decade, the study of the nucleon structure entered a new era with the 
advent of GPDs which unify within the same framework form factors, parton 
distributions, helicity and transversity distributions. DVCS appears as the 
golden chanel to access GPDs, showing an early scaling behaviour supported by 
recent JLab experimental results. The lack of experimental data does not yet 
allow for a precise and model-independent extraction of these quantities from 
experimental observables. The future experimental programs at JLab 12 GeV and
CERN are expected to overcome this situation, and hopefully provide a 
model independent determination of the contribution of the quark angular 
momentum to the spin of the nucleon.

\ack

This work was supported in part by the U.S. Department of Energy (DOE) contract 
DOE-AC05-06OR23177 under which the Jefferson Science Associates, LLC, operates 
the Thomas Jefferson National Accelerator Facility, the National Science 
Foundation, the French Atomic Energy Commission and National Center of 
Scientific Research, and the GDR n$^{\circ}$3034 Physique du Nucl\'eon.

\end{document}